\documentclass[journal]{IEEEtran}

\usepackage[ruled]{algorithm2e}
\usepackage{graphicx}
\usepackage{url}
\usepackage{subfigure}
\usepackage{color}
\usepackage{booktabs}
\usepackage{fontawesome}
\usepackage{amsmath}
\usepackage{multicol} 
\usepackage{multirow} 
\begin{document}

\title{Multi-Protocol Location Forwarding (MPLF) for Space Routing}
%
%
%

\author{
        Xiangtong~Wang,
        Zhiyun~Jiang,
        Menglong~Yang, 
        Songchen~Han,
        and~Wei~Li
\thanks{
        Wei~Li was with the School of Aeronautics and Astronautics, Sichuan University, Chengdu,
China, e-mail: (see li.wei@scu.edu.cn).}
}

\markboth{Journal of \LaTeX\ Class Files,~Vol.~14, No.~8, August~2015}%
{Shell \MakeLowercase{\textit{et al.}}: Bare Demo of IEEEtran.cls for IEEE Journals}

\maketitle

\begin{abstract}
            The structure and routing architecture design is critical for achieving low latency and high capacity in future LEO space networks (SNs). 
            Existing studies mainly focus on topologies of space networks, but there is a lack of analysis on constellation structures, which can greatly affect network performance.
            In addition, some routing architectures are designed for networks with a small number of network nodes such as Iridium while they introduce significant network overhead for high-density networks (i.e., mega-constellation networks containing thousands of satellites).
            In this paper, we conduct the quantitatively study on the design of network structure and routing architecture in space.
            The high density, high dynamics, and large scale nature of emerging Space Networks (SNs) pose significant challenges, such as unstable routing paths, low network reachability, high latency, and large jitter.
            To alleviate the above challenges, we design the structure of space network to maximum the connectivity through wisely adjusting the inter-plane inter satellite link.
            we further propose Multi-Protocol Location Forwarding (MPLF), a distributed routing architecture, targeting at minimizing the propagation latency with a distributed, convergence-free routing paradigm, while keeping routing stable and maximum the path diversity.
           
            Comprehensive experiments are conducted on a customized platform \textit{Space Networking Kits} (SNK) which demonstrate that our solution can outperform existing related schemes by about 14\% reduction of propagation latency and 66\% reduction of hops-count on average, while sustaining a high path diversity with only $O(1)$ time complexity.

\end{abstract}
    \begin{IEEEkeywords}
        Space networks, Networking structure design, Multi-Protocol Location Forwarding, Routing Scheme.
   
        \end{IEEEkeywords}

\IEEEpeerreviewmaketitle

\newcommand\fig{Fig.}
\newcommand\tab{Tab.}
\newcommand\alg{Algorithm.}

\section{Introduction}
\IEEEPARstart{L}{ow} Earth orbit (LEO) mega-constellations, comprising thousands of satellites, have been proposed to offer global broadband service and have attracted much attention in recent years.
“NewSpace” companies plan to launch mega-constellations of hundreds to thousands of communication satellites into LEO in the coming years.
 Their proposals have already received the first stage of regulatory approval: SpaceX\cite{starlink,starlink2}, OneWeb\cite{oneweb}, and Telesat\cite{telesat} have obtained RF spectrum from the FCC for their constellations.

\begin{figure}[t!]
    \begin{center}
        \includegraphics[width=1\linewidth]{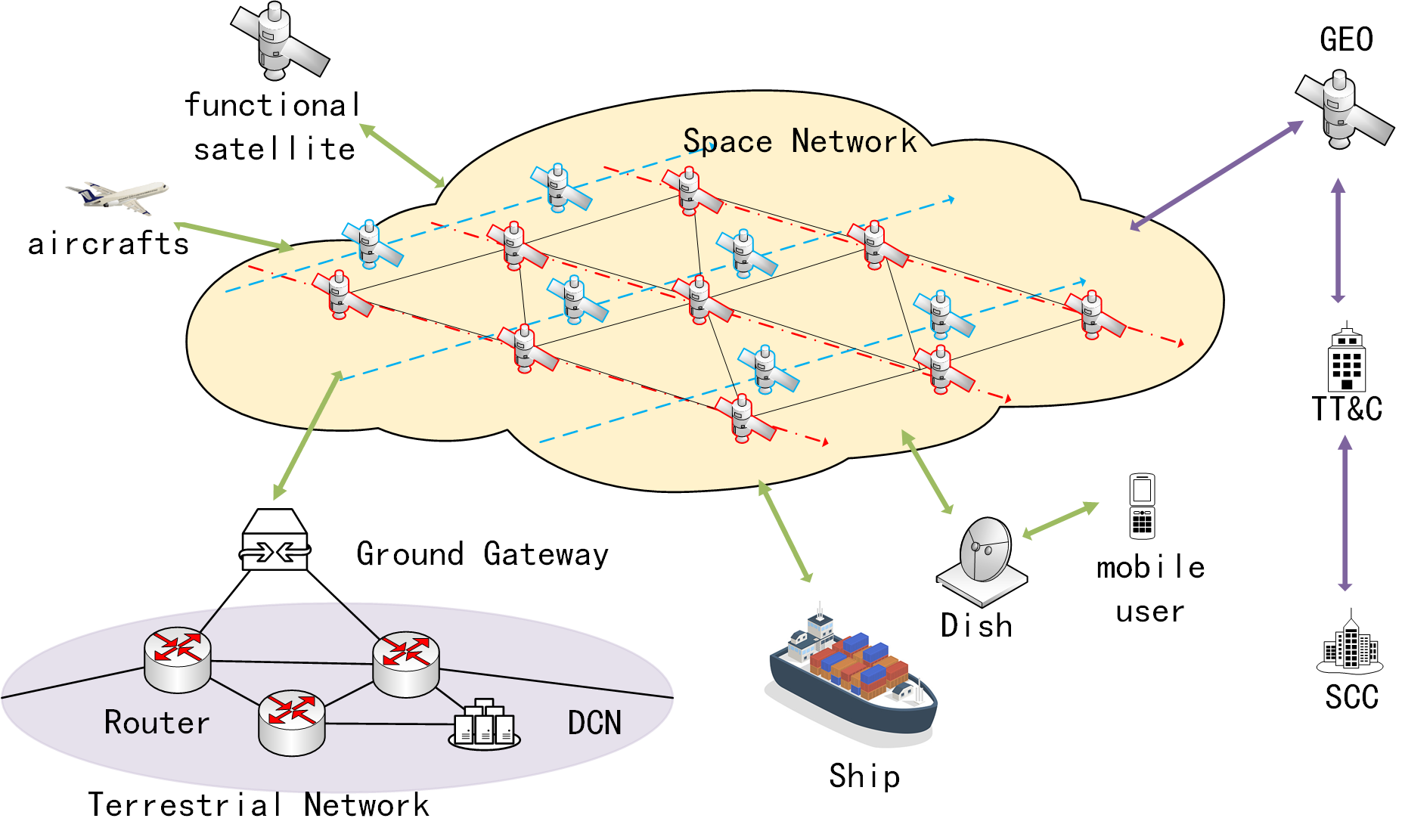}
    \end{center}
    \caption{Space Network Architecture.} 
        \label{fig:teaser}
 \end{figure}

\fig\ref{fig:teaser} shows the quintessential scenarios and use cases of future Space Network (SN). 
Such system could have a deep and lasting impact on the current Internet by extending its coverage to under-served communities, 
and providing lower latency over longer distances than present-day fiber. The SN could potentially serve as a core network and integrate into current terrestrial networks (TNs), 
and supporting the data transmission from functional satellite such as remote sensing satellite, weather observation satellites, etc.

However, realizing the full potential of these systems requires addressing new challenges posed by high-density large-scale and highly dynamics in the network, which bring a significant obstacle for datagram forwarding at the network layer.
Network layer protocols in TNs such as IP and ATM have a basic premise: datagrams can be forwarded according to statically addressed nodes for end-to-end reachability.
Even in wireless sensor networks (WSNs) such as Ad Hoc, nodes move around each other but still maintain a static topology, generating significant overhead to update the network information in each node\cite {clausen2003rfc3626,perkins2003rfc3561}, which is difficult to apply in SNs due to their density and scale.
More importantly, there are significant obstacles for addressing these challenges: the lack of network analysis tools that incorporate the dynamic behavior of LEO SNs\cite{kassing2020exploring}. 
This creates a substantial risk that research in this area may not keep peace with the rapid advances being made in the industry.

To address the above challenges, we explore how different configurations can affect the structure of the SN.
Based on the analysis, we propose the Multi-Protocol Location Forwarding (MPLF) architecture which employs a geographic based forwarding policy to enable end-to-end transmission between arbitrary terminals worldwide in a distributed and convergence-free manner over a dense constellation network.
Considering the unique dynamics of LEO SN, we developed a flexible and scalable simulator \textit{SNK} with the goal of quickly designing constellations and its network structure, obtaining reliable node information in dynamic networks, and visualizing routing paths.

Comprehensive experiments under Ground-Stations and Mobile Stations (MS) in various constellations conducted on SNK  to verify the effectiveness of MPLF. The results of experiments shows that compared with the other related algorithms, MPLF can significantly improve the overall network performance, by reducing about 14\% average propagation latency, 60\% jitter in maximum and 66\% average hops.

    
    
    

This paper is organized as follows.
Section \ref{sec:background} first provides a short overview about LEO/MEO satellite systems and ISL topology characteristics.
We formulate the SN model for simple description in Section\ref{sec:model}.
In Section \ref{sec:net}, the detailed analysis is given to explore how different configurations affect the structure of the SN.
The heart of the paper is then formed by Section \ref{sec:mplf} providing a detailed presentation of a new dynamic routing architecture MPLF for overcoming end-to-end problem in a density and dynamic SN.
In Section ref{sec:exp}, we present the performance analysis of MPLF and in Section \ref{sec:con} we conclude the paper.


\section{Background and Related Works}
\label{sec:background}

\subsection{Constellation and Networking Structure Design}

With the advent of multi-satellite systems, i.e., constellations, the traditional orbit design principles focused on a single satellite need to be extended to consider the performance of the entire constellation.
Some constellations \cite{iridium,starlink2,kuiper,telesat} have implemented inter-satellite link (ISL)  for inter-satellite information transfer, at which point the system can be abstracted as a network system, in which the design of space network structure must take into account the motion and traffic characteristics between satellite nodes in the system.

Satellite link establishment does not necessarily follow the principle of closest proximity, and some researchers have studied the constellation topology design, also called Link Assignment, which abstracts the FSA with a finite state machine and uses an optimization method to dynamically adjust the link establishment between satellites according to the load situation \cite{chang1995topological, chang1998fsa}, and the link can be established by satellites within any satellite visual range (LoS).
Bhattacherjee \cite{bhattacherjee2019network} ignored the nearest neighbor principle and proposed a topology design method for the giant constellation network and abstracted it as an integer linear programming problem, using moduli units to reduce the problem complexity for optimization. They also introduce several connectivity metrics for network evaluation.
Although not using the nearest neighbor principle can effectively increase the network throughput and reduce the average hop count of flows, etc., it is more like an extension of the study that mimics the network topology design in highly controllable DCNs \cite{al2008scalable,singla2012jellyfish} and lacks practicality, especially in highly uncontrollable environments in space where it is difficult to apply practically. because it would present additional technical hurdles that are difficult to overcome.
For example, the network overhead of link dynamic adjustment, which can be extremely burdensome when the constellation size rises, and is therefore only suitable for use in constellations with a small number of satellites and interstellar information transmission such as GNSS systems \cite{yan2020integer}.

Although some constellations have planned to implement satellite Internet without using ISL \cite{oneweb}, to achieve relay transmission in the form of Bent-pipe and ground stations, but it is difficult to achieve the performance of ISL system in terms of time delay, throughput, and robustness \cite{bhattacherjee2021towards}.
Therefore, establishing persistent links between neighboring satellites within and between orbits, respectively, is a more practical approach.
The persistent links need to consider the link construction pattern, which generally allows each satellite to establish persistent links with four or six adjacent satellites in a nearest neighbor manner, forming a local topology of Manhattan (+Grid) or triangular Manhattan pattern (*Gird)\cite{wood2001internetworking}. 
However, few works focus on the network structure affected by the configuration of constellation, specifically each satellite is connected to the neighboring satellites, but without considering which phase is connected.
The topology of space networks built on the Manhattan or triangular Manhattan pattern is relatively simple, but in the Walker configuration constellation there is a 'seam' phenomenon, which leads the 'last hop' ambiguity of some geographic information routing algorithms appear dead-end problem \cite{henderson2000distributed,tsunoda2006geographical}.

The structure of space network constructed based on the Manhattan or triangular Manhattan pattern is relatively simple, but there are "seams" in the Walker constellation, leading to dead-end problems in the "last-hop" ambiguity of some geographical routing algorithms \cite{henderson2000distributed,tsunoda2006geographical}, which we will discuss in Section \ref{sec:last_gap}.

\subsection{Routing in Space Networks}
Routing schemes have various criteria \cite{besta2020high} and can be categorized by:
Centralized and decentralized according to the computational entity;
connection-oriented type and connectionless according to whether or not to establish a state before data routing;
Topology-based or position-based routing according to whether or not to apply a location service;
Load-sensitive and load-sluggish according to load feedback;
static or dynamic according to the route change cycle.

Early satellite network services, such as Iridium \cite{iridium}, started with connection-oriented services, where packets between arbitrary source and destination nodes were forwarded along the same path, while connectionless forwarding was forwarded along different paths over time.
\cite{mauger1997qos,werner1997dynamic} proposes QoS-guaranteed routing methods for ATM networks, and provides shortest paths for ISL topologies abstracted as mesh networks. 
A similar connection-oriented routing technique is MPLS, a widely used in ISP networks. \cite{donner2004mpls} proposes MPLS for satellite networks using the architecture of Celestri\cite{celestri} as the network topology, and explores the possibility of its use in satellite networks.
The connection-oriented routing approach is only suitable for networks with few nodes, and providing centralized computation for networks with many nodes will have a significant routing overhead and therefore is not ideal for large-scale constellations.

To cope with the non-adaptability of centralized computational routing in satellite networks, some efforts\cite{ekici2000datagram,ekici2001distributed} have designed a distributed, connectionless routing algorithm with an Iridium-like constellation as the topology that can provide a path with a minimum propagation delay. 
Its algorithm abstracts the constellation topology as a two-dimensional mesh network, discretizes geographic addresses into logical addresses assigned to each satellite, and generates optimal forwarding paths hop-by-hop based on the relative distance from each node to its destination.

To overcome the dynamic nature of the network, some have proposed to achieve logical network statics through virtual nodes or virtual topologies\cite{chang1998fsa,jia2018routing,zhang2020application}.
This approach exploits the constellation topology's predictability, discretizes the dynamic topology according to the time slice, and assumes that the topology is constant within a specific time slice, thus generating routes using methods such as Dijkstra. 
Such methods shorten the adequate time of the static topology when the constellation size is large, thus requiring frequent updates of the topology, resulting in high overhead and not being suitable for large-scale constellations.

The virtual node approach\cite{mauger1997qos,ekici2000datagram,chen2019topology,korccak2009virtual} divides the geographical location into several cells, called virtual nodes, and assigns a fixed logical address to establish a mapping with the overhead coverage satellites. 
When the satellites cover the cell change, the following satellite inherits the logical address and network states, such as the routing table and other information of the previous satellite, maintaining the logical invariance  although the physical entity has already changed.
However, a large amount of bandwidth is required to exchange this information. The overhead is extremely high if each satellite caches a large routing table for the rest of the satellites in a large-scale network.

Geographic routing (also called georouting or position-based routing) is a routing principle that relies on geographic position information \cite{cadger2012survey,mauve2001survey,navas1997geocast} and have been widely used in WSNs.
Greedy forwarding and face routing are two of the earliest geographic routing strategies and have together formed the basis of numerous subsequent approaches.
In greedy forwarding, there are different strategies a node can use to decide to which neighbor a given packet should forwarded such as MFR (most forward within r) \cite{takagi1984optimal}, NFP (nearest with forward progress) \cite{hou1986transmission}, compass routing \cite{kranakis1999compass}, RN (random nearest) \cite{nelson1984spatial}.
Face routing is derived from Compass Routing II where faces on a planar graph are traversed using a technique known as the ’right hand rule’ (sometimes left hand rule instead) in which the algorithm keeps track of all the times it crosses the line connecting the source to destination\cite{kuhn2008algorithmic,karp2000gpsr}.
The main advantage of face routing is that compared with greedy forwarding, it guarantees delivery but with inefficiency caused by detours, which is not negligible in SNs due to its large scale.
The first geographical routing scheme that applied into LEO SNs  is introduced by \cite{henderson2000distributed}, which pose a last-hop ambiguity.
Thus, a mobility management scheme\cite{tsunoda2004supporting,tsunoda2006geographical}  that focuses on solving the last-hop ambiguity of geographical routing schemes has been proposed.



\subsection{Sim/Emulation Platforms}
Due to the characteristics of large scale, high dynamics, and a large number of nodes of space networks, the existing simulation platform cannot meet our current simulation demands on space networking. 
The ideal space network simulation platform needs to integrate three core functions: 1)space dynamics simulation and visualization, 2) network simulation 3) The high-performance connection scheme between them.

Many network simulators, including OPNET, NS2/NS3, Mininet, etc., and space simulators, including STK, MATLAB, SaVi, GMAT, etc., are considered for the development of space network simulation platforms, and the key technology is the organic combination of the two, i.e., to enable efficient interaction between the two platforms with scalability and flexibility.

\cite{handley2018delay} developed a satellite networking visualization platform through Unity 3D, which provides impressive visualization effects and can visualize the path evolution and end-to-end routing delay. However, the packet-level network simulation is still unachievable due to the lacking of support in network simulation.
\cite{kassing2020exploring} build a simulation platform Hypatia by combing NS3 and Cesium. It is the first to use Cesium to achieve visualization on the Web side with high flexibility. NS3 also supports virtual networks built by users to perform packet-level simulations. 
Although the webside in Hypatia is underdeveloped and the component interaction is weak, it still provides a good idea. On the other hand, \cite{lai2020starperf} combines the powerful network simulator mininet with STK, which has obvious superiority in satellite simulation, and proposes the StarPerf simulation platform and uses STK/Connect as a middleware to connect them. However, the limited expandability in STK leads to the visualization side lacking flexibility in showing customized entities such as ISLs.

Although some space network simulators have been proposed, there is still a lack of unified platforms that can fully implement packet-level measurements and powerful visualization of space networks.
The visualization platforms that present packet processes like ONOS\cite{berde2014onos} are of great value for future network development to understand existing phenomena.
\section{Space Network Model}
\label{sec:model}

\textbf{\textit{Definition 1:}} To tackle the temporal variations in the SN,
we denote an ordered \textbf{time set} as $ \mathcal{T}= \{t_1, t_2,\cdots \}$
The elements of the set $\mathcal{T}$ are called time stamps, where $t_i \le t_{i+1}$.
The network topology that without tISLs can be considered unchanged between
adjacent time stamps. $t_{i+1}- t_i$ represents the minimum time granularity of scenario change, and the time granularity with each element represent one second of time can be simply described as a time set $\mathcal{T} = \{1, 2, T\}$. 

\textbf{\textit{Definition 2:}} The \textbf{network topology} at each time stamp
$t \in \mathcal{T}$ can be formulated as an undirected graph $\mathcal{G}^t = (\mathcal{V},\mathcal{E}^t )$, where $\mathcal{V}$ is the set of network nodes and $\mathcal{E}^t$ is the set of active communication links.

Specifically, assume $\mathcal{V}$ is the set of all network nodes such
that $\mathcal{V} = \mathcal{V}_S\cup   \mathcal{V}_G\cup  \mathcal{V}_M $ where
$\mathcal{V}_S = \{s_1,\cdots, s_i,\cdots,s_{\mathcal{N}_S}\}$ , 
 $\mathcal{V}_G = \{g_1,\cdots, g_i,\cdots,g_{\mathcal{N}_G}\}$ and 
 $\mathcal{V}_M = \{m_1,\cdots, m_i,\cdots,m_{\mathcal{N}_M}\}$ represent the node sets of satellites, Ground stations (GSes) and Mobile Terminals (MTs) respectively.
Similarly, $\mathcal{E}^t$ is the set of all types of links in the network and composed by Inter-Satellite Links (ISL), Groundstation Satellite Links (GSL) and Mobile terminal Satellite Links (MSL) which denoted as $\mathcal{E}^t = \mathcal{E}^t_{S\leftrightarrow S}\cup  \mathcal{E}^t_{S\leftrightarrow G}\cup  \mathcal{E}^t_{S\leftrightarrow M} $.
We also describe all vertexes and edges that adjacent to $v$ at time $t$ as $\mathcal{V}_{adj}^t(v)$ and $\mathcal{E}_{adj}^t(v)$ respectively.

\textbf{\textit{Definition 3:}} 
If the costs of GSLs $\mathcal{E}^t_{S\leftrightarrow G}$ are not considered, \textbf{Paths} between vertexes $\{g_{s}, g_{d} \} \subset \mathcal{V}_G $ at each time stamp $t \in \mathcal{T}$  can be formulated as a set $\mathcal{P}^t_{g_{s} \rightarrow g_{d}} = \{  p_1,\cdots,p_i,\cdots,p_{\mathcal{N}_P} \} $, where the each elements represents a oriented link set $p = \{\mathbf{e}_1,\cdots,\mathbf{e}_i,\cdots,\mathbf{e}_{\mathcal{N}_{p}} \}$.



\textbf{\textit{Definition 4:}}
As described in Walker\cite{walker1984satellite}, the \textbf{constellation} is generally described as $T/P/F$ where $T$ is the total number of satellites; $P$ is the number of equally spaced planes, and $F$ is the relative spacing between satellites in adjacent planes. The change for satellites in neighboring planes equals $F \times 360 / T$.
The value range of the phase factor is $F \in [0, P-1]$. At $F=F_m$, the phase difference between satellites in two adjacent orbits achieves maximum where:
\begin{equation}
    \label{eq6}
   F_{m}=\left\{
    \begin{aligned}
    P/2 -1& , & \text{P~is~even.} \\
    (P-1)/2 & , &\text{P~is~odd.}
    \end{aligned}
    \right.
    \end{equation}
    Note that the constellation formulates the same architecture as $F=0$ and $F=P$. 
In this paper we use $N/P/F$ to illustrate a constellation where the $N$ is the number of satellites per orbit.

We also introduce the bias set $\mathcal{B}$ of inter-orbit link to describe the structure of constellation, and it will be detailed discuss in \ref{sec:ISL}.

\section{Structure Design and analysis in space Network}
\label{sec:net}
\begin{table*}[htbp]
	\caption{Types of Link in space network.}
	\centering
	\scalebox{0.96}{
		\begin{tabular}{|ccc|c|c|c|}
			\hline
			\multicolumn{3}{|c|}{\multirow{2}{*}{Types of Link}}                        & \multirow{2}{*}{Connected Entities}                             & \multirow{2}{*}{Dynamic}  & \multicolumn{1}{l|}{\multirow{2}{*}{At Edge}}                       \\
			\multicolumn{3}{|c|}{}                                                     &                                                                 &                           & \multicolumn{1}{l|}{}                                               \\ \hline
			\multicolumn{1}{|c|}{\multirow{4}{*}{\begin{tabular}[c]{@{}c@{}}Inter Satellite Link\\ (ISL)\end{tabular}}}           & \multicolumn{1}{c|}{\multirow{2}{*}{\begin{tabular}[c]{@{}c@{}}Persistent\\ ISL\end{tabular}}} & \begin{tabular}[c]{@{}c@{}}intra-orbit ISL\\ (iISL)\end{tabular} & adjacent satellites in same orbit             & \faTimes & \faTimes \\ \cline{3-4}
			\multicolumn{1}{|c|}{}                                                     & \multicolumn{1}{c|}{}                                           & \begin{tabular}[c]{@{}c@{}}Inter-orbit ISL\\ (sISL)\end{tabular} & satellites in adjacent orbits                 & \faTimes & \faTimes \\ \cline{2-4}
			\multicolumn{1}{|c|}{}                                                     & \multicolumn{1}{c|}{\multirow{2}{*}{\begin{tabular}[c]{@{}c@{}}Temporary\\ ISL\end{tabular}}} & \begin{tabular}[c]{@{}c@{}}Encounter ISL\\ (eISL)\end{tabular} & satellites in crossed orbits                          & \faCheck & \faTimes \\ \cline{3-4}
			\multicolumn{1}{|c|}{}                                                     & \multicolumn{1}{c|}{}                                           & \begin{tabular}[c]{@{}c@{}}Inter-shell ISL\\ (oISL)\end{tabular} & satellites in adjacent shells                  & \faCheck & \faTimes \\ \cline{1-4}
			\multicolumn{3}{|c|}{\multirow{4}{*}{Ground station satellite link (GSL)}} & \multirow{2}{*}{NGEO satellite and ground station}              & \multirow{2}{*}{\faCheck} & \multirow{2}{*}{\faCheck}                                           \\
			\multicolumn{3}{|c|}{}                                                     &                                                                 &                           &                                                                     \\ \cline{4-4}
			\multicolumn{3}{|c|}{}                                                     & \multirow{2}{*}{GEO satellite and ground station}               & \multirow{2}{*}{\faTimes} & \multirow{2}{*}{\faTimes}                                           \\
			\multicolumn{3}{|c|}{}                                                     &                                                                 &                           &                                                                     \\ \cline{1-4}
			
			\multicolumn{3}{|c|}{\multirow{2}{*}{Mobile station satellite link (MSL)}} & \multirow{2}{*}{satellite and mobile station}                   & \multirow{2}{*}{\faCheck} & \multirow{2}{*}{\faCheck}                                           \\
			\multicolumn{3}{|c|}{}                                                     &                                                                 &                           &                                                                     \\ \hline
		\end{tabular}
	}
	\label{tab:ISL}
\end{table*}

The structure of satellite networks are more like a hybrid structure of the mesh, ring and tree topology, which is significantly different from terrestrial networks and inseparable from routing scheme design. 
In this section, we introduce all the types of links that make up a satellite network, which shown in \tab\ref{tab:ISL}.
\subsection{Inter Satellite Links (ISLs)}
\label{sec:ISL}
\textbf{Intra-orbit Links (iISL).}
Without taking orbital perturbations into account, each satellite follows the same direction and velocity in the same orbit (plane). If a satellite establishes a link only with a neighbor satellite in the same orbit the topology can be considered invariant, as this is the only case of intra-orbit link in this paper.

\textbf{Inter-orbit links (side links, sISL).}
\label{sec:sISL}
Inter orbit links, i.e., the side links, refer to links that are established between two satellites in the same direction of movement but in adjacent orbits. 
Due to the relative homogeneity of iISL, sISLs are the main reason affecting the overall topology of the network. The main factors affecting the construction of sISL include:
\begin{itemize}
    \item The constellation phase factor $F$.
    \item Number of sISL per satellite.
    \item Which satellites in adjacent orbits establish sISL with.
\end{itemize}
\begin{figure}[htbp]
    \begin{center}
        \includegraphics[width=1\linewidth]{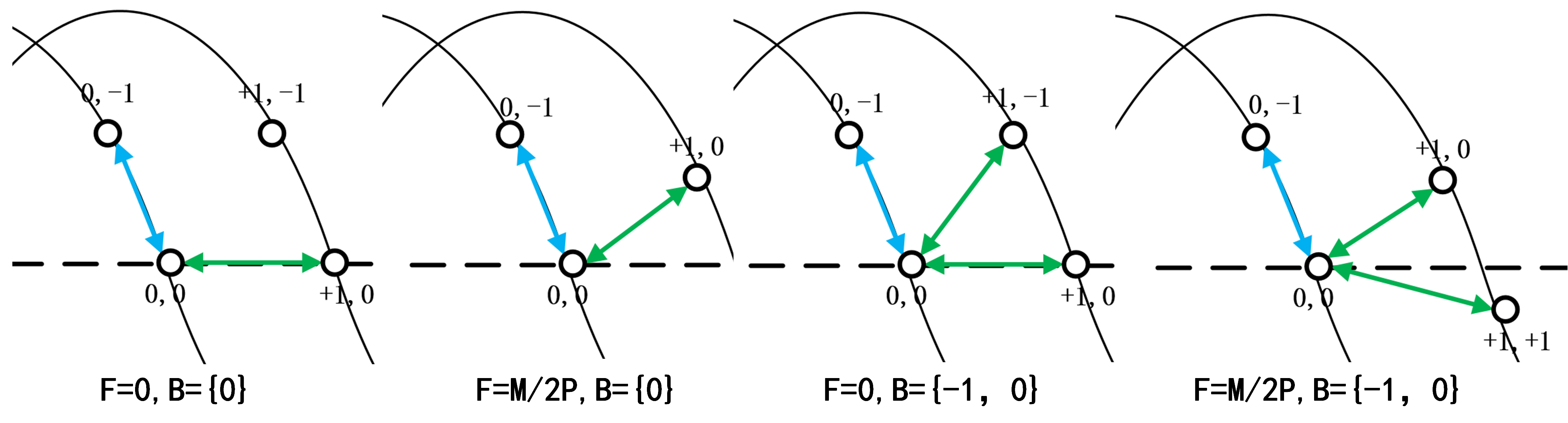}
    \end{center}
    \caption{The patterns of iISL (blue) and sISL (green) in +Grid (left two) and *Grid (right two). } 
        \label{fig:makeISLs}
 \end{figure}
\fig\ref{fig:makeISLs} shows four typical patterns of sISL building.
A satellite can build sISL with another in adjacent orbit with 0 bias($\mathcal{B}=\{0\}$), -1 bias ($\mathcal{B}=\{-1\}$) or both ($\mathcal{B}=\{-1, 0\}$). 
Generally, we denote the pattern of 4 ISLs per satellites as +Grid \cite{bhattacherjee2019network} and 6 ISLs per satellite as *Grid.
The inter-orbit satellite link is generally a link established between the current satellite and a neighboring satellite in adjacent orbit.
Note that there is another case where the link is a radom encounter between two satellites, which we call a encounter ISL (eISL).

\begin{figure}[htbp]
    \begin{center}
        \includegraphics[width=1.0\linewidth]{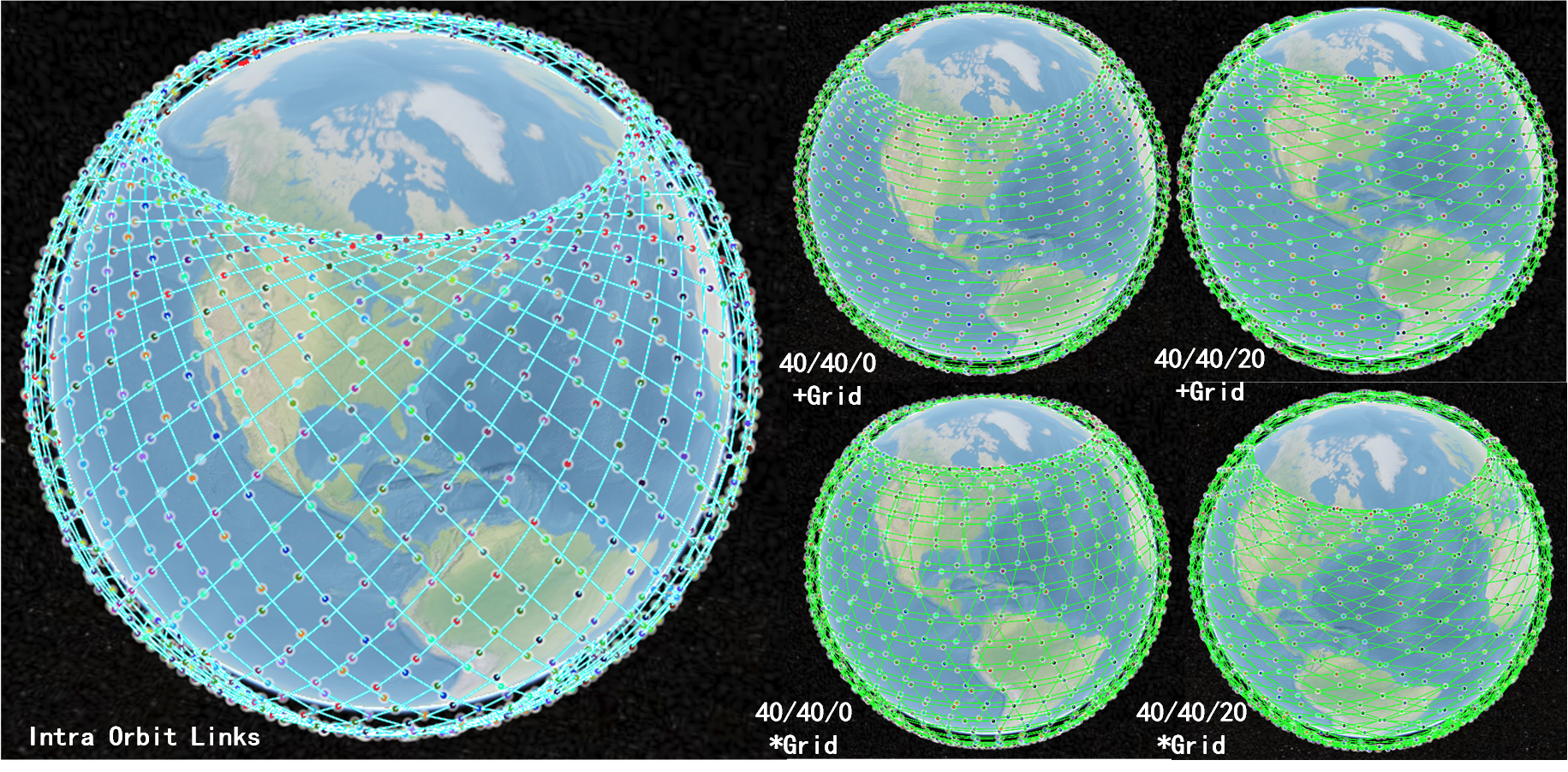}
    \end{center}
    \caption{Inter Satellite Links} 
        \label{fig:ISLs}
 \end{figure}

 For most LEO SNs, iISLs already provide sufficient SW-NE connectivity, while East-West connectivity and North-South connectivity require careful construction of sISLs.
 \fig\ref{fig:ISLs} shows a constellation with the size $N =40, P=40$ and various configuration where all iISLs (left cyan line) are identical, but sISLs (right green line) vary dramatically.
 It can be seen that the more uniform the connectivity of the constellation in all directions, the better performance of networking overall if the user demand is evenly distributed globally.

\textbf{Encounter ISL (eISL).}
\label{sec:eISL}
The networks generated from the above ISLs provide a mesh topology, but they are two different meshes in any one region. - One usually moves to the northeast (NE) and the other southeast (SE), also known as the 'seam'\cite{henderson2000distributed}, making it possible to establish a persistent ISL. There is no doubt that the link between them can provide the network great connectivity boost which we called as Encounter ISL (eISL).

\begin{figure}[htbp]
    \begin{center}
        \includegraphics[width=0.7\linewidth]{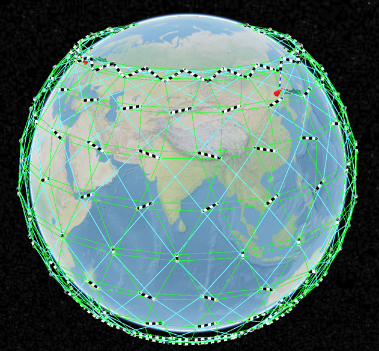}
    \end{center}
    \caption{Encounter ISL (Black and white) between NE direction and SE direction satellites. } 
        \label{fig:eISLs}
 \end{figure}
\begin{figure}[htbp]

 \subfigure[Duration of eISLs.]
 {
 \begin{minipage}{0.45\linewidth}
 \includegraphics[scale=0.35]{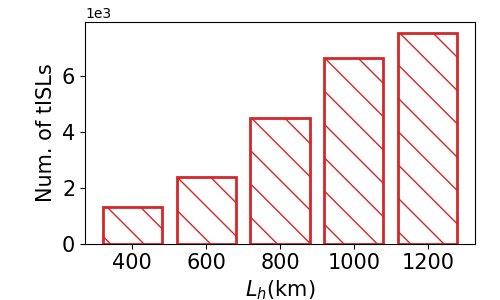}
 \end{minipage}
 }
 \subfigure[Num. of eISLs]
 {
 \begin{minipage}{0.45\linewidth}
 \includegraphics[scale=0.35]{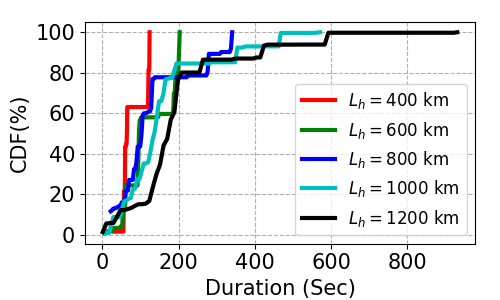}
 \end{minipage}
 }
    \caption{eISL between two orthogonal satellites.} 
            \label{fig:eISLs_cdf}
\end{figure}

The relative speed between the two satellites in eISL is high, about 5 to 12 km/s.  It is unclear for a satellite when, with which satellite, and for how long the link will be established.
 \fig\ref{fig:eISLs_cdf} (a) plots the number of eISLs and \fig\ref{fig:eISLs_cdf} (b) shows the duration of tISLs between two encountering satellites in orthogonal orbits under the constellation of 20/20/0, where $L_h$ is the distance threshold at which encountering satellites start to building eISL. We observed that over 80\% eISLs lasted less than 200s and the number of eISL increases significantly with increasing $L_h$.


\textbf{Inter-shell links (oISLs)}
Satellite networks can achieve full-time global coverage by building multi-layer (shell) constellations.
Different orbital height lead to different rotational velocities of satellites relative to the Earth between different shells, which makes it difficult to establish a persistent ISL between different shells.
However, the closer the height of orbit and the closer the velocities, the longer it takes to establish temporary links between satellites in adjacent shells.
Thus, although increasing the cost, stability can be effectively improved by increasing the number of shells, such as the strategy of using GEO as a controller in some SDN-based satellite networks \cite{bao2014opensan}.

Due to the complexity of the problem, eISL and oISL are not introduced in the latter part.

\subsection{Edge Links}
Edge links, including Ground station Satellite Links (GSLs) and Mobile Station Satellite Links (MSLs), are generally the links that connects forwarding satellites and earth or mobile stations, which are difficult to establish persistent links due to the high dynamics between satellites and station nodes.
The construction of the edge link is generally determined by the handover policy, in which the Ground Station (GS) dynamically selects the appropriate adjacent satellite among the coverage satellites, including maximum elevation angle (MEA), maximum service time (MST) or one-way latency\cite{zhang2022enabling}. However, it is simpler and more efficient to link all satellites together, which is the approach we take in this paper.
Since the coverage satellites can be considered as two mesh networks with overlapping but large topological distances, with a so-called 'seam' between them (\ref{sec:last_gap}), the GS should be connected to at least one of each SE and NE satellite for better connectivity in the absence of eISL.




\subsection{Connectivity Analysis}
The phase factor $F$ and the bias set $\mathcal{B}$ have an important impact on the construction of the sISLs, They affect the pattern of each satellite ISL and thus determine the global structure of the network, which can lead to ISLs with different directional distributions.

From \fig\ref{fig:makeISLs} we already know that, the sISLs in $F=0, \mathcal{B}=\{0\}$ pattern are parallel to the equator, which is called Horizontal Ring\cite{ekici2000datagram}, and it effectively increases the connectivity of the network in the East-West direction\cite{handley2018delay}. 
In fact, directional connectivity reflects the distribution of ISLs over different directions, i.e., the better the connectivity in a certain direction, the higher the proportion of ISLs in that angle, which will reduce the path zigzag in that direction and thus reduce the propagation latency.

\begin{figure}[htbp]
    \begin{center}
        \includegraphics[width=1.0\linewidth]{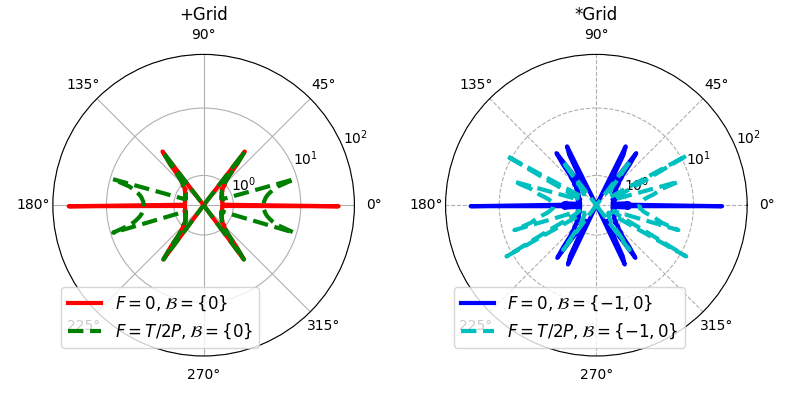}
    \end{center}
    \caption{$h(\alpha)$ in constellations with +Grid (a) and *Grid (b).} 
        \label{fig:polar}
 \end{figure}

In order to describe the directional connectivity of the constellation under different configurations, we define the $\alpha$ as the angle between the ISL and Equator plane as:
\begin{eqnarray}
\alpha = \frac{\pi}{2} - \arccos{ (\frac{\mathbf{e}\cdot \mathbf{e}_z}{\|\mathbf{e}\|\cdot \|\mathbf{e}_z\|})}
\end{eqnarray}
where the $e_z$ is the normal vector of Equator plane, i.e., the rotation axis of Earth.
We count the probability density function $h(\alpha)$ during time $\mathcal{T}$ as follows, which is shown in \fig\ref{fig:polar} at a polar coordination. 
\begin{eqnarray}
    h^\mathcal{T}(\alpha) = \frac{\sum\limits_{\mathcal{T}}h^{t}(\alpha) }{\|\mathcal{T}\|}
\end{eqnarray}

  We observe that
  iISL consists of orbits with $53^\circ$ inclination, providing sufficient SW-NE and NW-SE connectivity in all patterns.
  However, only the sISL in $F=0,\mathcal{B} = \{-1,0\}$ (solid blue line) provides additional ISL at an inclination of about $63^\circ$, which improves the North-South connectivity and thus reduces the zigzag connectivity, compared to the other patterns.

\section{Routing architecture in space}
\label{sec:mplf}

In this section, we bring the concepts of the Multi-protocol Label Switching (MPLS) \cite{rosen2001rfc3031} based network routing scheme and its expansion on the satellite network. 

In addition, we often refer to some interrelated concepts in routing\cite{tanenbaum2003computer,besta2020high,peterson2007computer} such as Switching and Forwarding, indicating the action to take when a packet arrives.
The former represents querying the packet egress port by standard matching and sending the data, while the latter means getting the packet egress port by fuzzy matching, i.e., there is a calculation step before sending the data.
Therefore, we use forwarding instead of switching in this paper.


\subsection{MPLS Concepts for NGSO Constellation Networking}

The basic idea of MPLS is very simple: at the ingress point of an MPLS network, the ingress label edge router (LER), packets are classified according to the information carried in the packet (e.g., source/destination address, class of service, etc. ) or network-related information (e.g., ingress port), or a combination of them.
A group of packets processed in the same way is called a Forwarding Equivalence Class (FEC). It is also possible to bundle a group of FECs and use a label for that union. This process is called aggregation.
Then, a unique locally valid fixed-length label is selected for packets belonging to a particular FEC and attached to each packet. A subsequent Label Switching Router (LSR) examines the packet's label, replaces it with the new label already specified, and forwards the packet to the next LSR based on the information stored in the table until it reaches the egress point of the network (egress LER) . The unidirectional path of packets through the MPLS domain is called the Label Switched Path (LSP).

MPLS has been proposed for geostationary orbit (GEO) systems with multiple point beams, primarily as a way to simplify the complexity of on-board switching \cite{ors2001providing}. However
in LEO satellite network systems, functional elements of MPLS, such as ingress LERs, egress LERs, or LSRs, must be mapped to physical entities of the space network, such as satellites or GSes\cite{donner2004mpls}.
Routing and rerouting of paths is of key interest because it affects the routing computation workload and routing performance. Therefore, the connection-oriented routing approach also faces many challenges in the highly dynamic topology between labeled edge routers (GSes) and labeled switching routers (Satellites), where the label distribution protocol (LPD) is affected. This is the motivation for us to proposed connectionless Multi-Protocol Location Forwarding method.

\subsection{Multi-Protocol Location Forwarding (MPLF) Architecture}



    

\begin{figure}[htbp]
    \centering

    \subfigure[MPLF between the data link layer and network layer.]
    {
    \begin{minipage}{1\linewidth}
    \centering
    \includegraphics[scale=0.73]{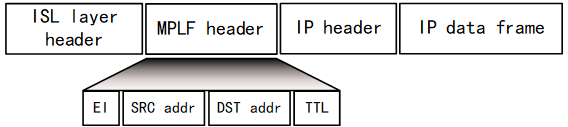}
    \vspace{5pt}
    \end{minipage}
    }

    \subfigure[Illustration of MPLF.]
    {
    \begin{minipage}{1\linewidth}
    \centering
    \includegraphics[scale=0.24]{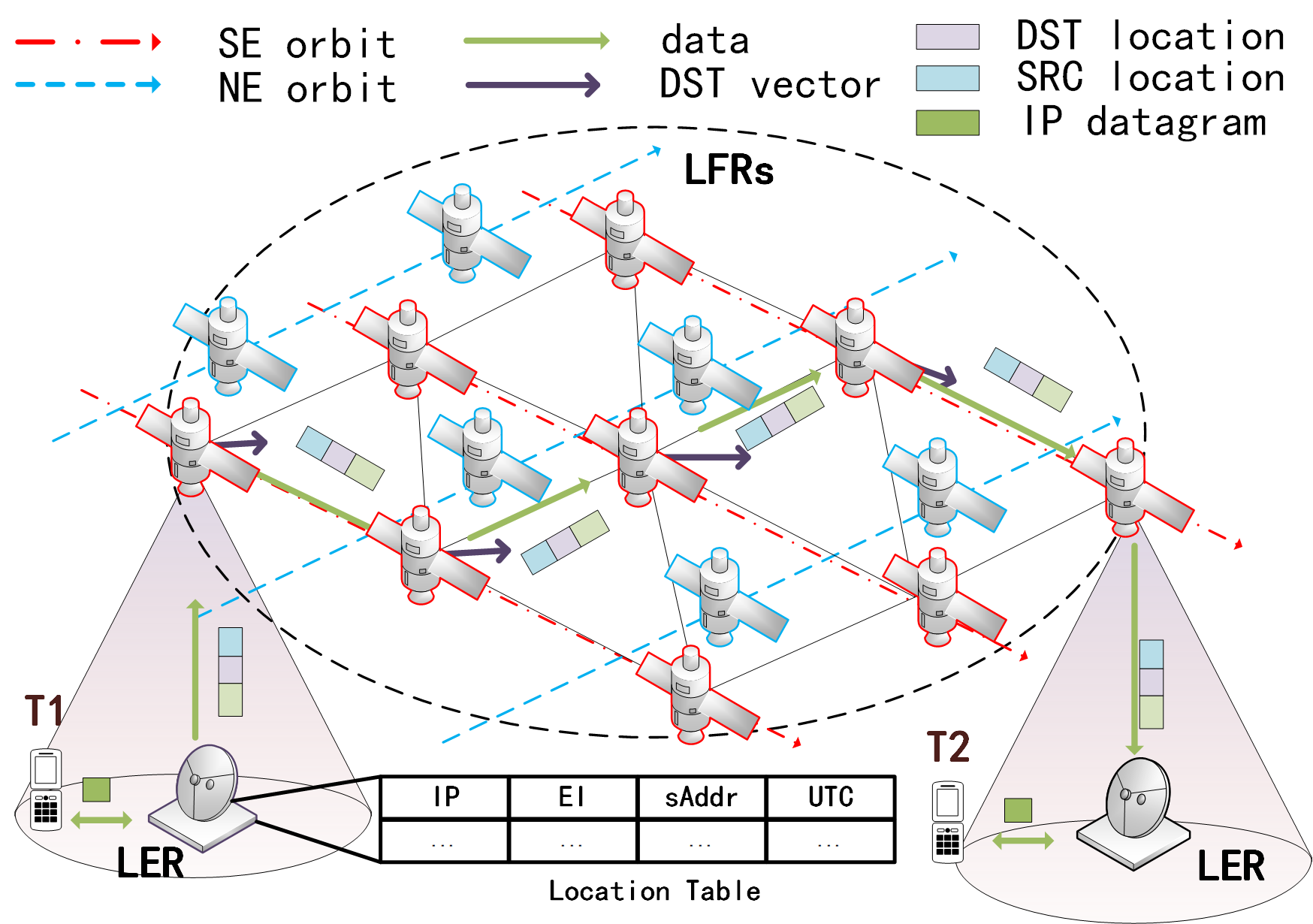}
    \vspace{5pt}
    \end{minipage}
    }

    \caption{MPLF architecture.}
    \label{fig:mplf}
    
\end{figure}

When two hosts communicate through the space network via LERs, datagrams pass through different satellite groups due to the changing location of the forwarding satellites, which poses a challenge to static network addresses.
Based on the concept of MPLS, we propose a connectionless routing architecture called Multiprotocol Location Forwarding (MPLF), which lies between the network and link layers (see \fig\ref{fig:mplf} (a)) and uses a distributed datagram forwarding method to compute the next hop using satellite location and port direction information.
Some of the definitions in MPLF are as follows.

\begin{itemize}
    \item \textbf{Location Forwarding Router (LFR).} The satellite in the network could be considered as LFR and it forward the packets according to information in packet's header and self-location to the next LFR, until the egress point (egress LER) is reached.
    \item \textbf{Location Edge Router (LER).} The ground station (GS) and mobile station (MS) both could be considered as LER, which to encapsulate high-level packets with constructed header and send them to associated satellites. Besides, the LER needs to convert the address of the data, which we will describe in Section\ref{sec:coord}.
    \item \textbf{Edge Location Server (ELS)}. LER can not store too large location tables in local due to the limited storage resources, so an ELS similar to Domain Name Server (DNS) can be built and connected with a terrestrial network, making it possible to query ELS through the terrestrial network to get the location of the destination LER when the local location table cannot be queried.
    \item \textbf{Equipment Identifier (EI).} An identifier that uniquely represents the device.
\end{itemize}
We equivalently use LFR and switch satellite to describe the same entities, and similarly for LER and MT or LER and GS.
To achieve MPLF, we assume that each satellite that we consider a Location Forwarding Router (LFR) must know and can update the position of itself and the adjacent satellite that exists in ISL in real-time, which is easy to achieve relying on existing GNSS or LEO GNSS \cite{reid2018broadband}.
The MPLF procedure that transmit IP packet from T1 to T2 is depicted in \fig\ref{fig:mplf}(b), which as follows.
\begin{itemize}

    \item [1)] T1 first sends an IP packet with destination T2 to source LER (i.e.,the ingress LER) $g_s$.

    \item [2)] Source LER $g_s$ queries the Equipment Identifier (EI) and space address (sAddr, Cartesian coordination format) of destination LER (i.e., the egress LER) $g_d$ by location table, packet the MPLF header with IP datagram, and send it to associated LFR $s_i$.

    \item [3)] The LFR $s_i$ receives the packet, takes out the EI from the MPLF header, queries the association table.
    
    \item [4)] If the EI does not exist in the association table, LFR calculates the port of next hop based on forwarding strategies .
    \item [5)] If EI exist in the association table, sends the datagram to the destination LER via corresponding channel.
    \item [6)] If the calculated next hop is the last hop, then drop the datagram for achieving loop-free.

\end{itemize}
After the destination LER $g_{d}$ receives the datagram, it save the source coordination $\mathbf{x}(g_{s})$ with the UTC time stamp to the local Location Table for updating. 
MPLF does not require the implementation of complex label distribution protocols like RSVP-TE\cite{awduche2001rsvp} or CR-LDP \cite{jamoussi2002constraint} at LERs, its only package the source and destination address to the datagram with coordination conversion. It is a connectionless routing method, and there is another calculation step before the general 'match-action' paradigm. 
In summary, the MPLF architecture has the following features and advantages:
    \begin{itemize}
        \item \textbf{Convergence-free.} Only the current FLR location and destination location are considered for each forwarding, and no forwarding table exists or needs to be updated.
        \item \textbf{Mobility management is not required.} Since there is no forwarding table, there is also no need for mobility management for networks with topology changes.
        \item \textbf{Support for multiple protocols} Similar to MPLS, MPLF is between the network layer and the link layer and can support multiple protocols.
        \item \textbf{Loop-free.} When packets have the same incoming and outgoing ports, they will be dropped by the node.
        \item \textbf{Support for multi-layer SNs.} The addressing system uses three-dimensional coordinates (See \ref{sec:coord}) and therefore supports the transfer of packets between different orbital layers.
       
    \end{itemize}

    However, in order to implement such an architecture, there are some details that we will discuss later, including location forwarding strategies (Section \ref{sec:forward}), address translation between ground and space (Section \ref{sec:coord}), and dealing with 'cross-seam' in the Walker constellation Section \ref{sec:last_gap}.
\subsection{Forwarding strategies}
\label{sec:forward}

\begin{algorithm}
    \label{alg:cpi}
    \caption{MPLF-I with CPI strategy}
    \LinesNumbered 
    \KwIn{ last FLR $s_{in},~s_{in} \in \mathcal{V}_{adj}(s_i)$\\
        \ \ ECI Coordinations of \\
        \ \ \ \ destination LER $\mathbf{x}(s_d)$ \\
        \ \  \ \  current FLR $\mathbf{x}(s_i)$ \\
        \ \ \ \ adjacent FLRs $\mathbf{x}(s_{j}),~ s_j \in \mathcal{V}_{adj}(s_i) $
        }
    \KwOut{outport $\mathbf{e}_{next}$}

        $ \textit{cos}_{min} \leftarrow 0$\;
        $\mathbf{e}_{d}\left\langle s_i,s_d \right\rangle  =  \mathbf{x}(s_d)-\mathbf{x}(s_i)$ \;
        \For{ $s_j \in \mathcal{V}_{adj}(s_i) $  }{
            
            $\mathbf{e}(s_j) \leftarrow \mathbf{x}(s_{j}) - \mathbf{x}(s_i)$ \;

            $\textit{cos} = \mathbf{e}(s_j) \cdot \mathbf{e}_{d}\backslash ( \|\mathbf{e}(s_j)\|\|\mathbf{e}_{d}\| )$\;
            \If {$\textit{cos}_{min} \ge \textit{cos} $}{
                $\textit{cos}_{min} \leftarrow \textit{cos}$\;
                $\mathbf{e}_{next} = \mathbf{e}(s_j)$
            }
           
        }
    \end{algorithm}
\begin{algorithm}
    \label{alg:nfp}
    \caption{MPLF-II with NFP strategy}
    \LinesNumbered 
    \KwIn{ last FLR $s_{in},~s_{in} \in \mathcal{V}_{adj}(s_i)$\\
        \ \ ECI Coordinations of \\
        \ \ \ \ destination LER $\mathbf{x}(s_d)$ \\
        \ \  \ \  current FLR $\mathbf{x}(s_i)$ \\
        \ \ \ \ adjacent FLRs $\mathbf{x}(s_{j}),~ s_j \in \mathcal{V}_{adj}(s_i) $
        }
    \KwOut{outport $\mathbf{e}_{next}$}

        $d_{min} \leftarrow 0$ \;
        \For{ $s_j \in \mathcal{V}_{adj}(s_i) $  }{
            
            $d_j \leftarrow \| \mathbf{x}(s_{j}) - \mathbf{x}(s_i)\|$ \;

            \If {$d_{min} \ge d_j $}{
                $d_{min} \leftarrow d_j $\;
                $\mathbf{e}_{next} = \mathbf{e}(s_j)$
            }
           
        }
    \end{algorithm}

Based on the position information, the greedy forwarding strategies such as NFP (nearest forward progress), i.e., forward to the neighbor is closer to the destination\cite{hou1986transmission,henderson2000distributed} or CPI (compass routing I)\cite{kranakis1999compass}, i.e., forward to the neighbor with the closest slope to that of the line segment connecting the destination can be applied to MPLF. 
We refer to the MPLFs that implement the above two forwarding policies as MPLF-I and MPLF-II, respectively.
The corresponding forwarding algorithm is shown in \alg\ref{alg:cpi} and \alg\ref{alg:nfp}.


The LFR needs to query the local coordination as well as the address of neighboring LFR at each forwarding to determine the next hop LFR. 
Thus, a LRF needs to interact with its neighboring LFR at regular intervals to pass coordination information to each other.
In addition, although the LFR are fast and relatively far away from each other, they do not need extremely accurate coordination information, which means that the mutual coordination interaction does not actually need high frequency, and we set 1Hz in the simulation to transfer coordination information to each other.

\subsection{Coordination Conversion}
\label{sec:coord}

Since the Geographic coordinate system is not continuous in the $180^\circ W / 0^\circ E $, which can be ambiguous at NFP or CPI forwarding, we use Earth Centered Inertial (ECI) coordinate system  and Earth-centered, Earth-fixed (ECEF) coordinate system for datagram addressing.
ECEF is generally used for ground stationary or low-speed motion nodes such as Ground-stations (GSes) or Mobile Stations (MSes). For satellites, it is more appropriate to use the ECI coordinate system that takes into account the rotation of the Earth. The conversion between the two coordinates systems need to be linked to Coordinated Universal Time (UTC) \cite{zhu1994conversion}, i.e., the ECI coordinates of an object are time-varying, regardless of whether the object is moving or not. 
It is necessary to consider when and where to convert the coordination. 

\begin{figure}[htbp]
    \begin{center}
        \includegraphics[width=1\linewidth]{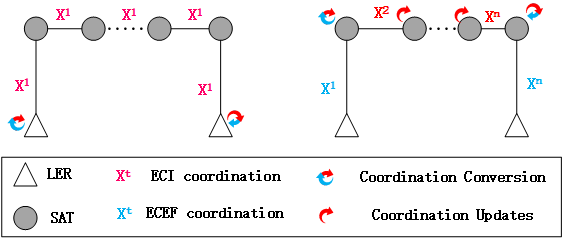}
    \end{center}
    \caption{Coordination Conversion in MPLF.} 
        \label{fig:conversion}
 \end{figure}


\fig\ref{fig:conversion} shows the two coordinate conversion strategies.
The first method is very simple, converting the ECEF coordinates of each LFR to ECI coordinates based on UTC time and updating them at each hop to obtain accurate coordinates. However this method is less efficient and can severely reduce the throughput of the LFR when the traffic becomes heavy.
Another method is to convert ECEF to ECI coordinates in LER only and always use the initial ECI coordinates in LFR. Although the coordinates will gradually increase the error with time, in the worst case, i.e., LER is located at the equator, if the Earth orbit speed is 27000km/s, the error is only 0.46m/ms, which is much lower than the coverage radius of the satellite. Therefore, the error performance is completely acceptable which is also the reason that we adopt this method.

\subsection{The Last Gap in MPLF}
\label{sec:last_gap}



\begin{figure}[htbp]
    \begin{center}
        \includegraphics[width=0.8\linewidth]{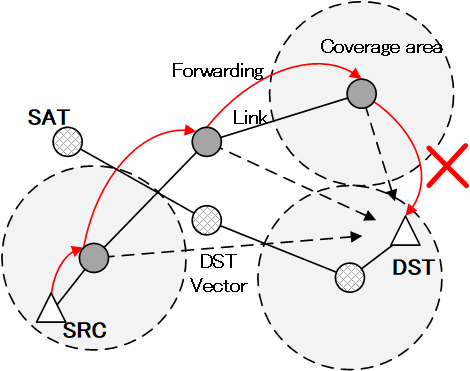}
    \end{center}
    \caption{Dead-end in CPI Forwarding.} 
        \label{fig:last-hop}
 \end{figure}

Compared with StarLink-like constellations, Iridium-like constellations have more prominent inclinations. They have two orbits with opposite rotations and large relative velocities, making it impossible to establish persistent ISL between the satellites across these orbits.
This situation creates the 'seam' between orbits and leads to some routing schemes to have the so-called 'dead-end' or 'last hop'\cite{henderson2000distributed,ekici2000datagram,roth2021implementation,tsunoda2006geographical}.
There is no guarantee that CPI and NFR can be delivered with certainty to the destination in networks with the 'seam'.
\fig\ref{fig:last-hop} shows the dead end which source node send packets through SE satellites (grey points) to the destination node that in the coverage of NE satellites (shadow points).
 However, it is foreseeable that as the number of satellites increases, the source node is able to deliver the packets to the destination node even the seam exists in the SN.
 To explore the impact of constellation density (i.e. the number of satellite in the SN) and latitude on packet delivery, we define reachability here for pairs $v_{s},v_{d}$ in paths $\mathcal{P}^t_{v_{s}\rightarrow v_{d}}$ at time $t$ as follows:

\begin{equation}
    \label{eq6}
    \psi ^t(v_{s},v_{d})=\left\{
    \begin{aligned}
    1,&~\text{if}~  v_{d} \in \mathcal{V}(e_{\mathcal{N}_p}),\exists e_{\mathcal{N}_p}  \in \mathcal{P}^t_{v_{s}\rightarrow v_{d}} \\
    0,&~\textit{other wise}.
    \end{aligned}
    \right.
    \end{equation}
where $\mathcal{V}(e_{\mathcal{N}_p})$ describe all vertex (i.e., two end of the link) linked by $\mathcal{N}_p$ and $\mathcal{P}^t_{v_{s}\rightarrow v_{d}}$ is paths set at time $t$. 

We further define the reachable probability $ p_r(\mathcal{G})$ to describe weather MPLF can find a path between two alternative vertexes $v_{s},v_{d}$ included by the satellite network $\mathcal{G}$ during $\mathcal{T}$, which is: 
 \begin{equation}
    \label{eq6}
    p_r(\mathcal{G})= \frac{\sum\limits_{\{v_{s},v_{d}\} }\sum\limits_{t=1}^{\mathcal{T}}\psi ^t(v_{s},v_{d})}{|\mathcal{T}|}, ~ \forall\{v_{s},v_{d}\} \subset \mathcal{V}
    \end{equation}

\begin{figure}[htbp]
    \begin{center}
        \includegraphics[width=1\linewidth]{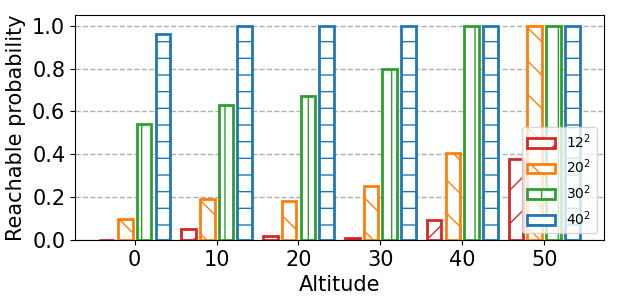}
    \end{center}
    \caption{Reachable probability among satellites and GSes in 40/40/0(f+0) constellation.} 
        \label{fig:HitRatio}
 \end{figure}

\fig\ref{fig:HitRatio} shows the  $p_r$ between GSes in SN with different latitudes.
We observe that with the increase of GS latitudes, the higher reachable probability it achieved due to the denser satellites. 
The constellation with $40^2$ size ($N=40,P=40$) maintains full reachable probability even at the lowest latitudes region.
The results show that with a dense LEO SN, $100\%$ reachable probability can still be achieved even the GSes located at the Equator that with the sparsest density of satellites.
Thus the effect of seams can be neglected in such constellations.

\subsection{Multipathing in MPLF}
\label{sec:mpr}

LEO Constellation can provide multiple paths between a certain end-to-end pairs and achieve load balancing through multi-pathing mechanisms such as Equal-Cost Multipathing (ECMP) \cite{hopps2000analysis}.
A GS $g$ may associate multiple satellites $\mathcal{V}_{adj}(g)$ at time $t$ by GSLs $\mathcal{E}_{S\leftrightarrow G}^t(g)$.
we focus on paths from $\mathcal{V}_{adj}(g_s)$ to $\mathcal{V}_{adj}(g_d)$ given by the routing algorithm as the set of equal-cost paths $\mathcal{P}$.
Thus, the number of paths from $g_s$ to $g_d$ obeys $|\mathcal{P}_{g_s \rightarrow g_d}| \leq |\mathcal{V}_{adj}(g_s)|\ast |\mathcal{V}_{adj}(g_d)|$ where $|\mathcal{V}_{adj}(g)|$ is the number of satellites that GS $g$ associated.

However, when only few different links between multiple paths exist, the paths are too correlated and less independent, which making it difficult to share the load equally.
In graph theory, the average degree is defined as $d_{avg} =2 n_e/n_v$, where $n_v,n_e$ are the number of vertexes and edges in a direction graph, and $1 \leq d_{avg} \le n_v$ \cite{west2001introduction}.
To better characterize the potential performance of multipathing over different paths generated by routing algorithm, we take the $\gamma $ as the metrics to describe the independency of paths as follows:
\begin{eqnarray}
    \gamma (\mathcal{P})= \frac{2}{d_{avg}}=\frac{ n_v}{n_e}= \frac{|\mathcal{E}_{S\leftrightarrow S}(\mathcal{P})| }{|\mathcal{V}_S(\mathcal{P})|}
\end{eqnarray}
where $|\mathcal{E}^t_{S\leftrightarrow S}(\mathcal{P})|$ is the number of ISLs in $\mathcal{P}$ and $|\mathcal{V}^t_S(\mathcal{P})|$ is the number of satellites (LFR) in $\mathcal{P}$.

\subsection{Complexity of MPLF}

Traditional graphical algorithms such as the Dijkstra and Bellman-Fold algorithms used in routing algorithms such as Distance Vector (DV) or Link State (LS) usually have a high complexity, i.e. $O(n^2)$ of Dijkstra and $O(n^3)$ of Bellman-Fold in worst case, where $n$ stands for the number of nodes in the network.
In addition, when small changes in the network occur, such as the attendance of eISL or changes in link values caused by satellite moving, creating significant overhead.
Therefore, the conventional shortest path algorithm have scalability problem, which is not suitable for the networks with large number of nodes.
In contrast, proposed MPLF provides minimum propagation latency paths with the complexity $O(1)$, and it takes a very short time to process the packets and the time is independent of the network size. Besides, it does not need to consider the effect of topology and link length changes to determine the next hop.


Another important point is the storage complexity of the MPLF. 
On one hand, given a specific satellite network, the forwarding port for next hop of each satellite is computed in real-time and can be embedded into the routing code before deploying the satellites. Hence, no additional space for routing tables is needed. 
On the other hand, all the other shortest path algorithms need at least a connectivity matrix, which is of size $P^2 \times N^2$ where $P$ is the number of planes and $N$ is the number of satellites in a plane. Furthermore, in order LER to process the incoming packets faster, a routing table of size may be necessary. Therefore, our algorithm has much less space complexity than the other schemes. 


\section{Simulation Study}
\label{sec:exp}




In oder to show the effectiveness of our proposed MPLF, we compare the paths generated by MPLF that combined with two forwarding strategies and denoted as MPLF-I (NFP) and MPLF-II (CRI), with the Shortest Paths (SP) and Least Hop (LH) paths generated by Bellman-Ford\cite{kershenbaum1993telecommunications}.
Large scale characteristics in space networks lead to dynamic changes between links and high propagation latency, which is the major part of link cost calculation. Therefore, in experiments we only take propagation latency into consideration.
For performance evaluation of the MPLF we conducted three main experiments.
\begin{itemize}
    \item \textbf{Experiment I:} We compare the one-way propagation latency of paths generated by MPLF with the paths generated by SP and LH among three cities pair.
    \ \item \textbf{Experiment II:} We discuss the diversity and evolution of paths generated by MPLF, SP and LH.
    \item \textbf{Experiment III:} We applied the MPLF to the terminals that moved over 20000km and compare the one-way latency of paths generated by MPLF with the paths generated by SP and LH.
\end{itemize}

To achieve high-fidelity experiments, we develope a simulation framework \textit{SNK} \cite{snk} for satellite networking, and all the experiments are conducted on it.
One of the component in this framework is \textit{visualizer}, a Cesium-based visualizer that displays entities (e.g., satellites, GSes and aerocrafts) and interacts (e.g., ISLs,GSLs) in real time in the browser.
For downloading and analysis the data from satellite networks, we also develope the \textit{visualizer-backend} to interact with \textit{visualizer}. All the entities in network such as satellites, ground station, mobile station, ISLs, etc. are built by the third component \textit{scenario} in \textit{SNK}.
In the experiments, we set orbit height to $550km$, inclination to $53^\circ$ and minimum satellite elevation to $40^\circ$.

\begin{figure*}[htbp]
    \centering

    \subfigure[Harbin to London.]
    {
    \begin{minipage}{0.33\linewidth}
    \centering
    \includegraphics[scale=0.48]{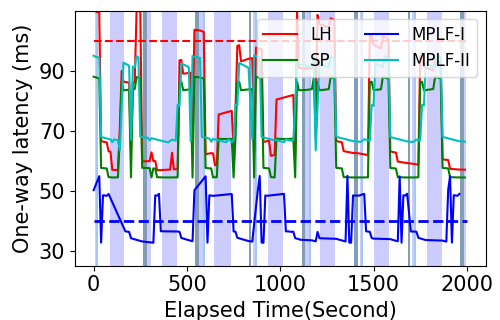}
    \end{minipage}
    }\subfigure[London to San Francisco]
    {
    \begin{minipage}{0.33\linewidth}
    \centering
    \includegraphics[scale=0.48]{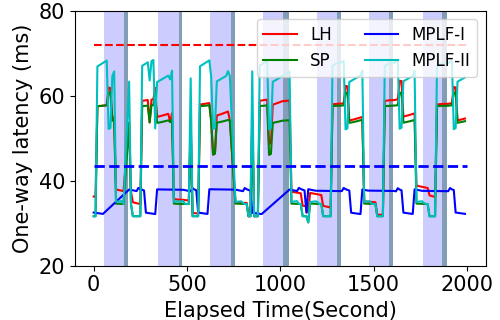}
    \end{minipage}
    }\subfigure[San Francisco to Shanghai]
    {
    \begin{minipage}{0.33\linewidth}
    \centering
    \includegraphics[scale=0.48]{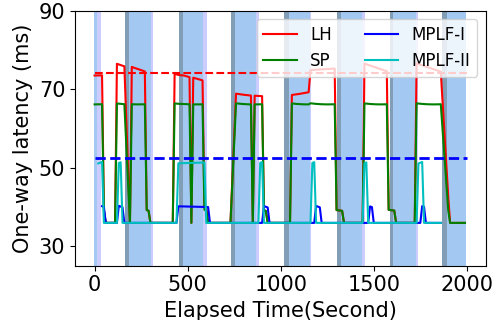}
    \end{minipage}
    }
    
    \caption{One way propagation latency between city pairs under 20/20/0(+Grid) constellation.}
    \label{fig:e2e_methods}
    
\end{figure*}

\begin{figure*}[t!]
    \centering

    \subfigure[Harbin to London.]
    {
    \begin{minipage}{0.33\linewidth}
    \centering
    \includegraphics[scale=0.48]{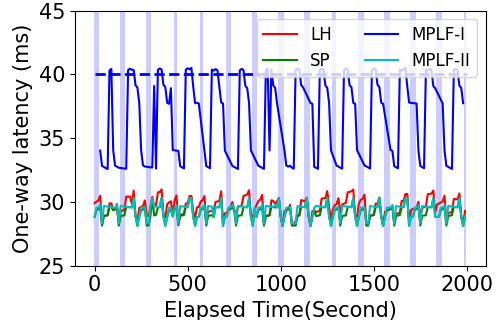}
    \end{minipage}
    }\subfigure[London to San Francisco]
    {
    \begin{minipage}{0.33\linewidth}
    \centering
    \includegraphics[scale=0.48]{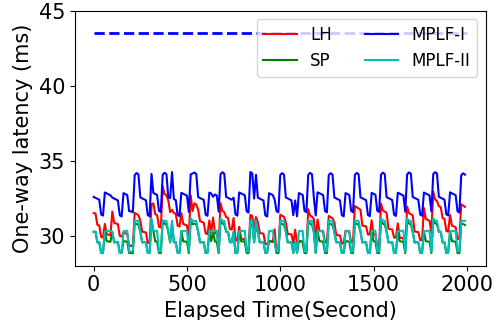}
    \end{minipage}
    }\subfigure[San Francisco to Shanghai]
    {
    \begin{minipage}{0.33\linewidth}
    \centering
    \includegraphics[scale=0.48]{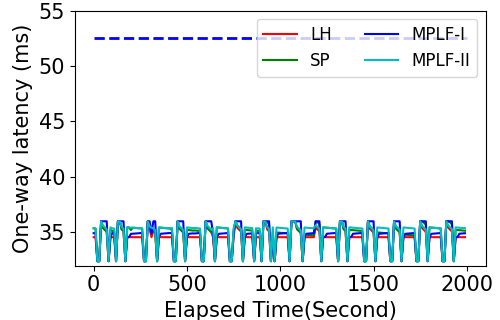}
    \end{minipage}
    }
    
    \caption{One-way propagation latency between city pairs under 40/40/0(*Grid) constellation.}
    \label{fig:e2e_methods2}
    
    \end{figure*}

    \begin{figure*}[t!]
        \centering

        \subfigure[Harbin to London.]
        {
        \begin{minipage}{0.33\linewidth}
        \centering
        \includegraphics[scale=0.48]{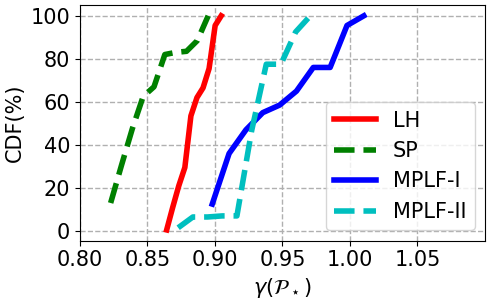}
        \end{minipage}
        }\subfigure[London to San Francisco]
        {
        \begin{minipage}{0.33\linewidth}
        \centering
        \includegraphics[scale=0.48]{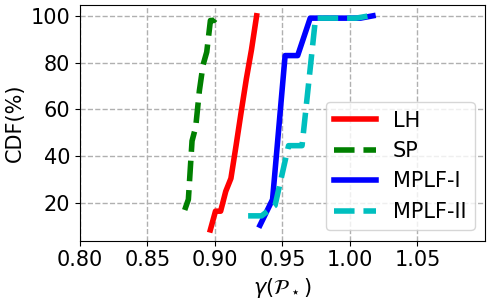}
        \end{minipage}
        }\subfigure[San Francisco to Shanghai]
        {
        \begin{minipage}{0.33\linewidth}
        \centering
        \includegraphics[scale=0.48]{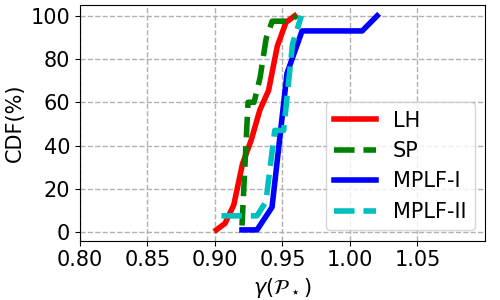}
        \end{minipage}
        }
        
        \caption{Path diversity between city pairs under 40/40/0(*Grid) constellation.}
        \label{fig:div}
        
    \end{figure*}

    \begin{figure*}[t!]
        \centering

        \subfigure[Harbin to London.]
        {
        \begin{minipage}{0.33\linewidth}
        \centering
        \includegraphics[scale=0.48]{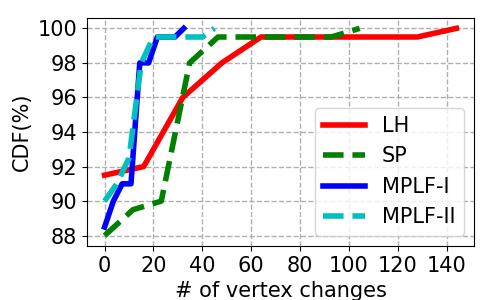}
        \end{minipage}
        }\subfigure[London to San Francisco]
        {
        \begin{minipage}{0.33\linewidth}
        \centering
        \includegraphics[scale=0.48]{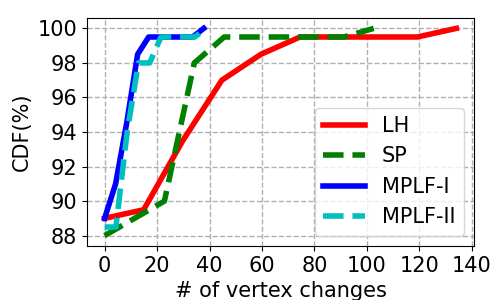}
        \end{minipage}
        }\subfigure[San Francisco to Shanghai]
        {
        \begin{minipage}{0.33\linewidth}
        \centering
        \includegraphics[scale=0.48]{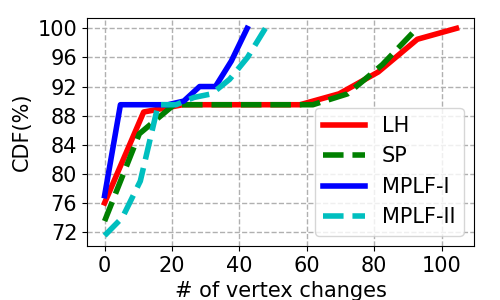}
        \end{minipage}
        }
        
        \caption{Path evolution between city pairs under 40/40/0(*Grid) constellation.}
        \label{fig:evo}
        
    \end{figure*}

\subsection{End to End latency in GSes}

We build constellations with the configuration $N/P/F = 20/20/0$ (i.e., with the size of $20^{2}$) and $N/P/F = 40/40/0$ (i.e., with the size of $40^{2}$) and both of them are patterned as +Grid pattern (See \ref{sec:sISL}).
We choose 3 typical cities pairs: Harbin-London, London-San Francisco and San Francisco - Shanghai at 4 different latitudes, as test connections for routing algorithms.
Simulation duration is set as $2000$ seconds and the time step is 10 seconds.
The one-way latency of Internet is drawn in red dash line which is given by WonderNetwork\footnote{https://wondernetwork.com/} and the geodesic latency (i.e. the time of geodesic length divided by $2c/3$ where c is light speed \cite{handley2018delay,kassing2020exploring}) is drawn in blue dash line.

\fig\ref{fig:e2e_methods} shows the One-way latency between city pairs in the constellation with $20^2$ size. 
Due to the absence of the accessing satellite for a GS, sometimes there is no valid path between cities. The red shadow is the invalid time of LH, the green shadow is SP's, the blue one is MPLF-I's and the cyan one is MPLF-II's. 
In \fig\ref{fig:e2e_methods} (a) and (b), MPLF-II has achieve a competitive result with distributed method comparing with LH and SP.
In \fig\ref{fig:e2e_methods}(c) the latency of MPLF-II is lower, however, the valid time of MPLF-II is also dramatically lower than LH and SP, the reason behind is that the reachable probability of MPLF-II is decreasing with the density of accessible satellites.
However, After improve the size of constellation from $20^2$ to $40^2$ (\fig\ref{fig:e2e_methods2}), the MPLF-II has achieved the best performance path in one way latency among cities which is almost indistinguishable from the path generated by SP. 
Also, if the number of satellites is large enough, the 'dead-end' problem in MPLF-II will not exist as there is no shadow in the figure. This clearly shows that our new proposed MPLF provides minimum one way propagation latency with the complexity $O(1)$.


\subsection{Diversity and Structure Evolution of Paths}
\label{sec:evo}

    
    
    
    

To avoid congestion or failures caused by a particular satellite, the SN can use path diversity in the routing scheme to provide multiple paths between the same GS pair.
With a routing algorithm, a satellite that associated by the source $g_s$ can at least find a path to the satellite associated with a destination $g_d$. The total number of paths between them is $|\mathcal{V}^t_{adj}(g)|\ast |\mathcal{V}^t_{adj}(g_d)|$ where $|\mathcal{V}_{adj}(g_s)|$ is the satellites that $g$ associated at time $t$.
We only take available paths in MPLF-II or MPLF-I into account and we observe that due to the 'dead-end' cases in MPLF-II and MPLF-I, the number of available paths of them is less than LH or SP.

\fig\ref{fig:div} shows the CDF of $\gamma$ in city pairs. 
We notice that MPLF-II and MPLF-I are get larger $\gamma$ comparing with LH/SP, which means it generates more dispersed paths and can achieve higher utilization in practice.
Over $60\%$ paths generated by SP get $\gamma = 0.85$, and paths generated by LH get $\gamma =0.88$ while the paths generated by MPLF-I and MPLF-II gets 0.93 and 0.95 respectively. The higher value of $\gamma$ represents higher utilization.
The main reason behind this behavior is that the node of paths generated by MPLF are more regularly distributed around a geodesic arc of SRC and DST nodes, and the utilized links are therefore more concentrated, which brings the concentrated $\gamma$.

\begin{figure}[htbp]
    \begin{center}
        \includegraphics[width=1.0\linewidth]{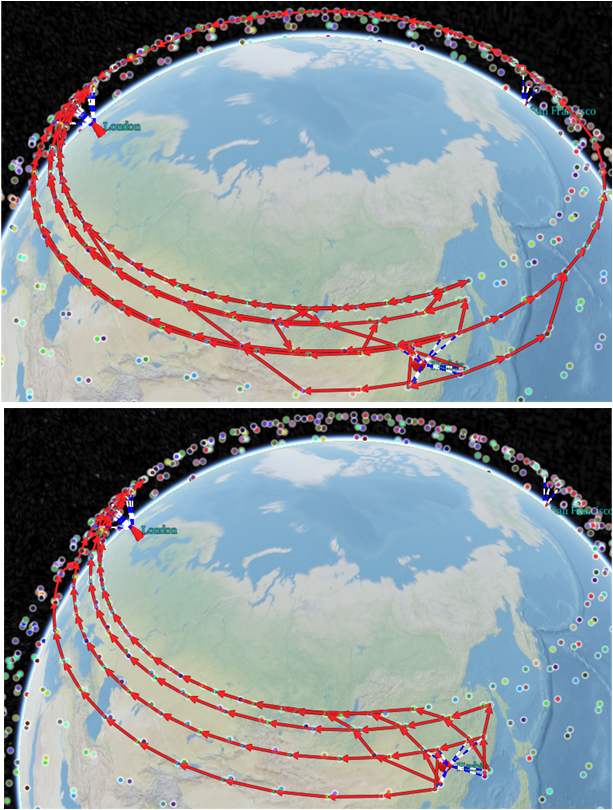}
    \end{center}
    \caption{All paths from Harbin to London that provided by LH/SP (upper) and MPLF (bottom) at an identical time. } 
        \label{fig:diversity}
 \end{figure}

In addition, as the latitudes of communicate GS is getting lower, the $\gamma$ is higher. we identify that such variations is caused by the gradually sparse satellites that GS associated
which reduce the probability of detours.
\fig\ref{fig:diversity} shows the paths $\mathcal{P}^t$ in SN. The left one is generated by LH and SP while the right one is generated by MPLF.


Besides the path diversity, we also examine the structure evolution of generated paths. For each connection, we measure the number of different vertexes (satellite) between the paths $\mathcal{P}^t$ and $\mathcal{P}^{t+1}$ over the adjacent time in simulation.
If the forwarding paths computed in two successive time-steps shows any different satellites, we count the number of different satellites as the vertex change, i.e., $\|\mathcal{P}^{t} \cup \mathcal{P}^{t+1} - \mathcal{P}^{t} \cap  \mathcal{P}^{t+1}\|$.
Across connections, we compute the CDF of these vertexes changes. 
\fig\ref{fig:evo} (a) and (b) show that in the median, over the 98\% paths generated by MPLF change less than 20 nodes in the connection of 'Harbin-London' and 'London-San Francisco' while LH and SP are 40.
In \fig\ref{fig:evo} (c), more than 88\% of the paths generated in all algorithms have a number of node changes below 20, but the rest of the paths generated by LH and SP are between 60 and 80, while MPLF is only between 20 and 40.

Unlike the current Internet, LEO network paths evolve rapidly, especially for the denser networks, with paths changing multiple times per minute. Changes in path forwarding nodes during routing can lead to high fluctuation in data transmission under network, so fewer changes in the number of nodes can allow for more robust end-to-end data transmission.

\begin{table*}[]
	\caption{Detailed information of moving LERs.}
	\label{tab:acs}
	\centering
	\scalebox{0.96}{
		\begin{tabular}{c|ccccccccccc}
			\toprule[2pt]
			\vspace{2pt}
			LER Name        & AC00     & AC01     & AC02     & AC03     & AC04     & AC05     & AC06     & AC07     & AC08     & AC09     & AC10     \\ \hline

			Start Point     & 0E,50N   & 18E,50N  & 36E,50N  & 54E,50N  & 72E,50N  & 90E,50N  & 0E,40N   & 0E,30N   & 0E,20N   & 0E,10N   & 0E,0N    \\
			End Point       & 180W,50S & 162W,50S & 144W,50S & 126W,50S & 108W,50S & 90E,50S  & 180W,40S & 180W,30S & 180W,20S & 180W,10S & 180W,0S  \\
			Distance (km)   & 19998.15 & 17461.25 & 15062.56 & 13020.83 & 11599.87 & 11081.69 & 19970.32 & 19972.31 & 19977.69 & 19984.97 & 19992.30 \\
			Velocity (km/s) & 10.1     & 8.81     & 7.60     & 6.57     & 5.85     & 5.59     & 10.08    & 10.08    & 10.08    & 10.08    & 10.08    \\
			\toprule[2pt]
			\multicolumn{8}{l}{\small $\bullet$ denotes the Distance is LER moving distance instead of end-to-end distance.}                         \\
		\end{tabular}
	}

\end{table*}

\begin{figure*}[htbp]
    \begin{center}
        \includegraphics[width=1\linewidth]{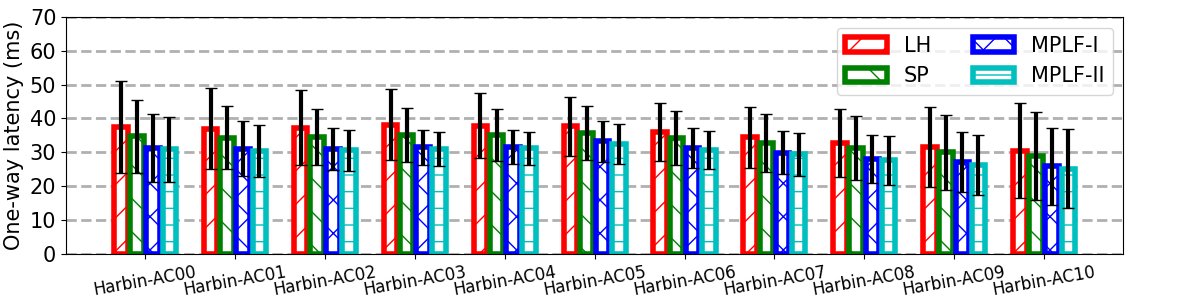}
    \end{center}
    \caption{The trajectory of moving nodes.} 
        \label{fig:acs_bar}
 \end{figure*}

 \subsection{Latencies between Fixed GS and Moving LER}


 \begin{figure}[t!]
    \centering

    \subfigure[Distribution of Avg. hop-counts of paths.]
    {
    \begin{minipage}{0.47\linewidth}
    \centering
    \includegraphics[scale=0.44]{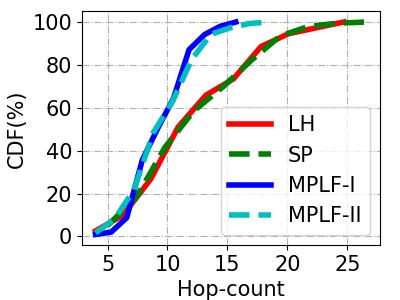}
    \end{minipage}
    }\subfigure[Distribution of paths stretches.]
    {
    \begin{minipage}{0.45\linewidth}
    \centering
    \includegraphics[scale=0.44]{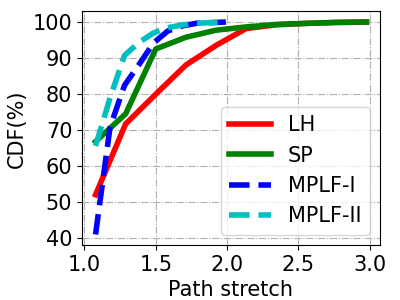}
    \end{minipage}
    }
      
    \caption{One way propagation latency among cities in 40/40/0(+Grid) constellation.}
    \label{fig:stretches_ac}
    
\end{figure}

The space network can also provide communication to mobile objects such as aircraft, ship or remote sensing satellite, therefore we wonder how the propagation latency of the moving target behaves. 
In order to demonstrate the propagation latencies and their distribution due to the terminals mobility, we consider a scenario with multiple aircraft (moving LER) at different starting points and end points from West to East.
We use the constellation $N/P/F = 40/40/0$ with +Grid pattern as the network.
The simulation duration is set as 1980 seconds and the time step is 10 seconds.
All the moving details of LERs are shown in \tab\ref{tab:acs}.

\textbf{Latency and jitter.} We randomly select the paths between Harbin and Moving LER generated by SP, LH, MPLF-I and MPLF-II, and plot their one-way propagation latency on average, minimum, and maximum under the same constellation, as shown in \fig\ref{fig:acs_bar}.
We observe that MPLF performs better results among all the LER pairs, reducing the average latency by about 5ms to 10ms compared with LH and SP, outperforming the other algorithms with about 14\%. 
Besides, the path fluctuations generated by MPLF are narrower than those generated by LH and SP in most connections. For example, the fluctuations of 'Harbin-AC03' and 'Harbin-AC04' under LH and SP are more than 20 ms, while it is only 8 ms under MPLF, about 60\% reduction, which can significantly reduce network jitter.






\textbf{Hops-count.} \fig\ref{fig:stretches_ac}(a) shows the distribution of hop-count in paths to reflect the one-way process latency and transmission latency in networking. We observe that the MPLF outperforms the LH and SP with about 66\% reduction in maximum hops and 50\% in hops jitter.

\begin{figure*}[htbp]
    \begin{center}
        \includegraphics[width=1\linewidth]{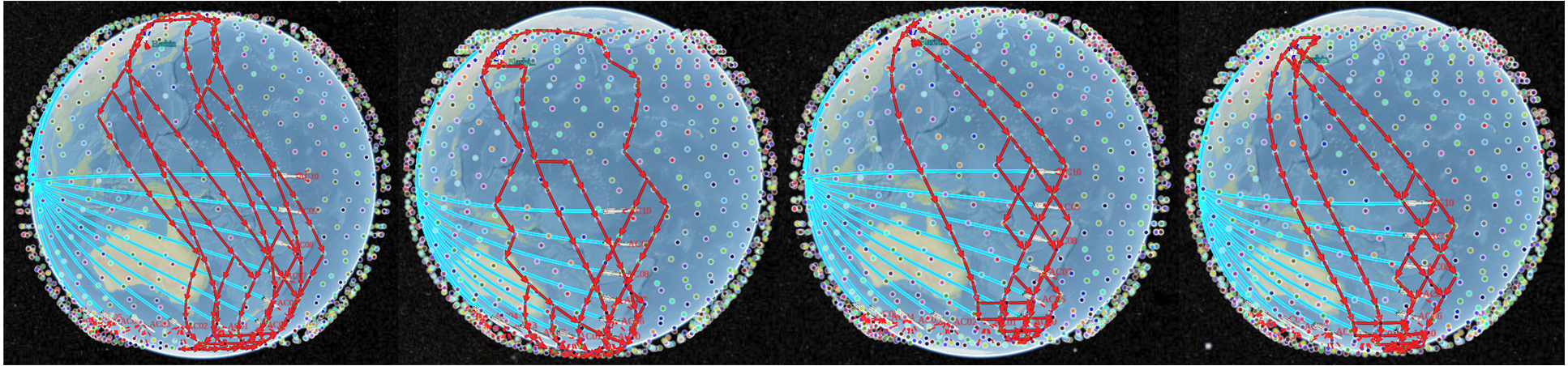}
    \end{center}
    \caption{The trajectory of moving nodes.} 
        \label{fig:acs}
 \end{figure*}

\textbf{Stretches.} Since the distances between pairs varying, we refer \cite{bhattacherjee2019network,kassing2020exploring} to quantify latency in terms of \textit{stretch}, i.e., the ratio of the path distance across the designed network and the geodesic distance. The lower stretch represents the less detours of the path.
\fig\ref{fig:stretches_ac}(b) shows the distribution of path stretch changes between 11 moving LER pairs. We observe that all the stretch are under 1.6 in the paths generated by MPLF while all the stretches are under 2.1 in the paths generated by SP and LH.
\fig\ref{fig:acs} shows the paths generated by LH,SP,MPLF-I and MPLF-II between Harbin and moving LER at the end point. The blue lines are the trajectories of moving LER.

From experiments results above, MPLF achieves better performance in terms of latency compared with LH and SP. 
The path diversity and structure evolution also illustrate that the MPLF can provides higher utilization.
The advantage of MPLF is not obvious when the range is extended to more GSes, but when applied to mobile terminals, MPLF gets the best results compared to other algorithms.
It is worth noting that MPLF is a distributed algorithm, and satellites also do not need to save global topological adjacencies. The efficiency of its algorithm, which has only $O(1)$ complexity compared to $O(n^2)$ or $O(n^3)$ of Dijkstra or Bellman-Ford, achieves better results than other methods, which is highly feasible in space.

\section{Conclusion}
\label{sec:con}
This paper studied the design of architecture to enable low-latency transmission in emerging LEO SNs.
We investigate the structure design of space networks and propose a distributed routing architecture MPLF, which incorporates the location and orientation of mobile satellites into next-hop decisions, but with dramatically lower complexity.
Comprehensive experiment results demonstrate that our algorithm outperforms related algorithms.

\ifCLASSOPTIONcaptionsoff
  \newpage
\fi

\bibliographystyle{IEEEtran}
\bibliography{ref}

\end{document}